\documentclass[openacc]{rstransa}

\newcommand*{\Ham}{{\cal{H}}}
\newcommand*{\HamB}{{\cal{H}}_B}

\newcommand*{\Heff}{\mbox{\boldmath$H$}}

\newcommand*{\SMM}{\mbox{\boldmath$S$}}
\newcommand*{\svec}{\mbox{\boldmath$\sigma$}}
\newcommand*{\tvec}{\mbox{\boldmath$\tau$}}
\newcommand*{\sx}{\sigma_x}
\newcommand*{\sy}{\sigma_y}
\newcommand*{\sz}{\sigma_z}

\newcommand*{\ex}{\mbox{\boldmath$e$}_x}
\newcommand*{\ey}{\mbox{\boldmath$e$}_y}
\newcommand*{\ez}{\mbox{\boldmath$e$}_z}
\newcommand*{\eS}{\mbox{\boldmath$e$}_S}
\newcommand*{\vecmu}{\mbox{\boldmath$\mu$}}
\newcommand*{\dvec}{\mbox{\boldmath$d$}}
\newcommand*{\gvec}{\mbox{\boldmath$g$}}
\newcommand*{\fvec}{\mbox{\boldmath$f$}}
\newcommand*{\jvec}{\mbox{\boldmath$j$}}
\newcommand*{\jsvec}{\mbox{\boldmath$j$}}

\newcommand*{\tauvec}{\mbox{\boldmath$\tau$}}

\begin{document}

\title{Non-equilibrium charge and spin transport in SFS point contacts}

\author{
C. Holmqvist$^{1}$,  W. Belzig$^{2}$, and M. Fogelstr\"om$^{3}$}

\address{
$^{1}$Department of Physics, Norwegian University of Science and Technology, NO-7491 Trondheim, Norway\\
$^{2}$Fachbereich Physik, Universit\"at Konstanz, D-78457 Konstanz, Germany\\
$^{3}$Department of Microtechnology and Nanoscience - MC2, Chalmers University of Technology,
SE-412 96 G\"oteborg, Sweden}
\subject{Physics}


\corres{Mikael Fogelstr\"om\\
\email{mikael.fogelstrom@chalmers.se}}

\begin{abstract}
The conventional Josephson effect may be modified by introducing spin-active scattering in the interface-layer
of the junction. 
Here, we 
discuss a Josephson junction consisting of two s-wave superconducting leads coupled over a classical spin that precesses with the Larmor frequency due to an external magnetic field.
This magnetically active interface results in a time-dependent boundary condition with different tunnelling 
amplitudes for spin-up and -down quasiparticles and where the precession produces spin-flip scattering processes. 
As a result, the Andreev states develop sidebands and a non-equilibrium population that depend on the details of the spin precession.
The Andreev states carry a steady-state Josephson charge current and a time-dependent spin current,
whose current-phase relations could be 
used for characterising the precessing spin. 
The spin current is supported by spin-triplet correlations induced by the spin precession and creates a feed-back effect on the classical spin in the form of a torque that shifts the precession frequency.

By applying a bias voltage, the Josephson frequency adds another complexity to the situation and may create resonances together with the Larmor frequency. 
These Shapiro resonances are manifested as torques and are, under suitable conditions, able to reverse the direction of the classical spin in sub-nanosecond time.
Another characteristic feature is the subharmonic gap structure in the dc charge current displaying an even-odd effect that is attributable to precession-assisted multiple Andreev reflections.

\end{abstract}


\maketitle

\section{Introduction}
Interesting spin phenomena may occur when ferromagnets are combined with superconductors (see \cite{eschrig2011} and \cite{linder2015} and references therein). Cooper pairs in 
a conventional superconductor have spin-singlet pairing which, if the superconductor is interfaced with a ferromagnet, extend into the ferromagnet. However, 
the exchange field inside the ferromagnet tries to align the two spins of the Cooper pairs and hence breaks the Cooper pairs apart resulting in a rapid decay 
of the superconducting correlations inside the ferromagnet. For the same reasons, the critical current of a Josephson junction with a ferromagnetic layer
sandwiched between the two superconductors decays rapidly with increasing thickness of the ferromagnetic layer
\cite{bulaevskii1977,ryazanov2001,kontos2002,buzdin2005}. On the other hand, if weak ferromagnetic interfaces with magnetisation 
directions differing from the magnetisation direction of the ferromagnetic layer are inserted, the spin-singlet correlations may be transformed into 
spin-triplet correlations which can survive over a long range within the ferromagnet layer \cite{bergeret2001,bergeret2005,houzet2007,braude2007,eschrig2008}.
As a result of this non-collinear magnetisation of the ferromagnetic layer, the critical current decays similarly to a supercurrent in a non-magnetic metal with
increasing junction length \cite{keizer2006,khaire2010}. So far, the existence of spin-triplet correlations has been measured in this indirect way.
A more direct way of detecting the spin-triplet correlations would be to measure the effects of the spin on the triplet correlations, e.g. by using phenomena explored in conventional spintronics such as spin-transfer torques and other means for creating magnetisation dynamics effects or magnetisation switching.
There has been theoretical work done in this direction
\cite{konschelle2009,mai2011,cai2010}
using approaches  based on the Bogoliubov-de Gennes equations \cite{waintal2001,waintal2002,michelsen2008,linder2011,kulagina2014} and Green's function methods \cite{zhu2004,zhao2008,houzet2008,braude2008,yokoyama2009,shomali2011}
as well as some experimental work investigating the coupling between the dynamics of magnetic moments and Josephson currents \cite{petkovic2009,barnes2011}, but to our knowledge there has been no experimental investigation of the coupling between magnetisation dynamics and induced triplet correlations.
This is a crucial step in developing superconducting spintronics applications \cite{linder2015}.
In this article, we will review recent work on how magnetisation dynamics of a nanomagnet couple to the induced spin-triplet correlations associated with the charge and spin Josephson effects, and discuss how the dynamic interactions between the induced spin-triplet correlations and the nanomagnet lead to non-equilibrium transport properties that can be used to probe the induced triplet correlations directly.

\section{Quasiclassical model}

\begin{figure}[!b]
\centering\includegraphics[width=5in]{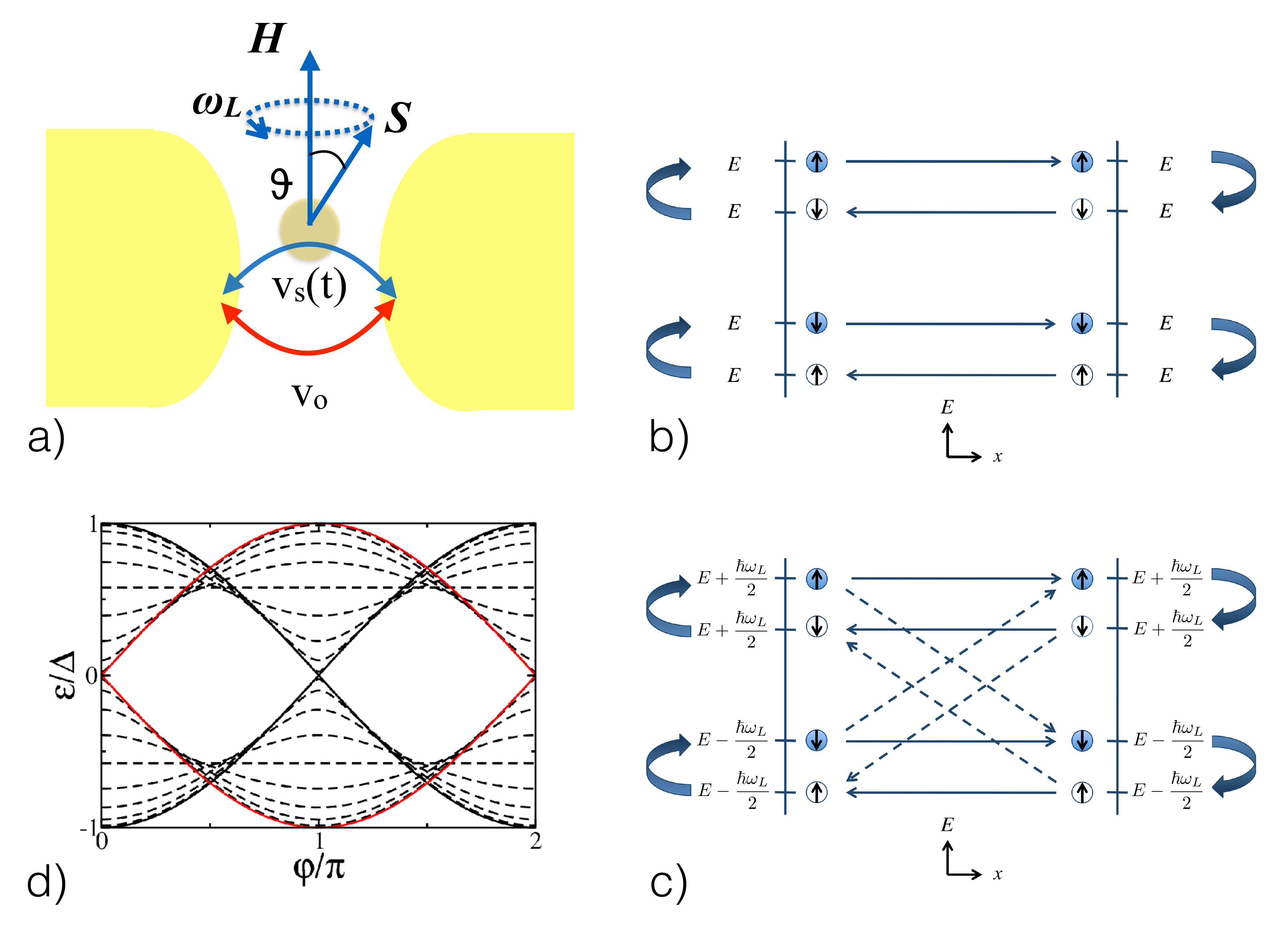}
\caption{(a) Two superconducting leads are coupled over the spin of a nanomagnet.
The tunnel junction is characterised by the hopping amplitudes $\rm{v_o}$ and ${\rm{v_s}}(t)$, where $\rm{v_o}$ is the spin-independent tunnelling and ${\rm{v_s}}(t)$ is the phenomenological time-dependent coupling generated by the nanomagnet, whose spin precesses with the frequency $\omega_L$ at the cone angle $\vartheta$.
(b) The schematics of conventional Andreev scattering between
two superconductors at phase difference $\varphi$. Constructive interference occurs at a phase-dependent energy $\varepsilon(\varphi)$ defining two energy-degenerate Andreev levels.
(c)
In addition to the spin-conserving tunnelling (solid lines),
the dynamics of the spin allows for tunneling processes with spin-flip scattering combined with an absorption or emission of the energy $\hbar \omega_L$ (dashed lines).
The combination of these tunnelling processes results in
a lifting of the spin-degeneracy of the Andreev levels in (b) and the appearance of time-dependent spin-triplet pairing amplitudes.
(d) For a junction
with a static spin, the Andreev-level spectrum's dependence on phase may be modified from a $0$ junction, 
${\rm{v_o}}>{\rm{v_s}}$, to a $\pi$ junction, ${\rm{v_o}}<{\rm{v_s}}$.
The black full line is $({\rm{v_o}},{\rm{v_s}})=(1,0)$ and the red line is $({\rm{v_o}},{\rm{v_s}})=(0,1)$. The dashed lines span between these two limits in increments
of $0.1$.}  
\label{fig1}
\end{figure}

Consider two ordinary BCS s-wave superconductors, with a phase difference $\varphi$, coupled over a nanomagnet as depicted in Fig.~\ref{fig1}(a). The nanomagnet may be a magnetic
molecule or a magnetic nanoparticle which we will treat as a classical spin, $\SMM$, with magnetic moment $\vecmu=\gamma\SMM$, and the
gyromagnetic ratio $\gamma$. The nanoparticle supports a few conduction channels when placed between the two metallic leads. 
If the nanomagnet is subjected to an external magnetic field, $\Heff$, it will precess when 
the effective field is applied at an angle, $\vartheta$, relative to the spin. 
$\Heff$ is an effective field that includes any r.f. fields needed to maintain precession, crystal anisotropy fields and demagnetisation effects.
The spin and the effective magnetic field couple via a Zeeman term,
$\HamB=-\gamma \SMM(t)\cdot\Heff$.
At finite tilt angle, $\vartheta$, the spin precesses with the Larmor frequency, $\omega_L=\gamma H$, where $H=\vert \Heff \vert$ is the magnitude of the effective field.
The spin dynamics are described by the Landau-Lifshitz-Gilbert equation of motion \cite{gilbert2004,tserkovnyak2005}
\begin{equation}\label{LLG}
\frac{d\SMM}{dt}=-\gamma \SMM(t) \times \Heff +\tvec(t),
\end{equation}
where the first term on the right-hand side is the torque produced by the effective field and the second term, $\tvec(t)$, is
a torque that collects effects caused by the
mutual coupling between the precessing nanomagnet and the superconducting quasiparticle system.

The coupling of the motion of the spin and the quasiparticle tunnelling over the spin enters via a time-dependent tunnelling term,
$\hat{\Ham}_T = \hat{\psi}_{L}^{\dagger} \hat{v}_{LR}(t) \hat{\psi}_{R}  +  H.C.,$
where $\hat{\psi}_{\alpha}$ is the usual spin-dependent Nambu-spinor that describes the superconducting state in lead $\alpha=\rm{R,L}$.  
The hopping matrix $\hat{v}_{LR}(t)\,(=\hat{v}^\dagger_{RL}(t)\equiv \hat{v}(t))$ has a spin-structure that may be parametrised into a spin-independent 
amplitude ${\rm{v_o}}$ and a spin-dependent amplitude ${\rm{v_s}}(t)$. It has the following matrix structure in the combined $4\times 4$ Nambu-spin space,
\begin{equation}\label{hopping element}
\hat{v}_{LR}(t)=
\left(\begin{array}{cc}
{\rm{v_o}}+{\rm{v_s}}(\eS(t)\cdot\!\svec)&0\\0&{\rm{v_o}}-{\rm{v_s}}\sigma_y(\eS(t)\cdot\!\svec)\sigma_y
\end{array}\right).
\end{equation} 
We use the time-dependent unit vector, $\eS(t)$, along $\SMM(t)=\vert\SMM\vert \eS(t)$ and include the magnitude $|\SMM|$ in the 
spin-dependent amplitude ${\rm{v_s}}$. Above, $\svec=(\sx,\sy,\sz)$ with $\sigma_i$ being the {\it i-th} Pauli matrix.
The spin-independent amplitude and the portion of the spin-matrix parallel to $\Heff$, 
${\rm{v_o}}+{\rm{v_s}}\cos \vartheta\,(\ez\!\cdot\!\svec)$, 
describe the tunnelling amplitudes for spin-up and spin-down quasiparticles,
while the portion perpendicular to $\Heff$,
${\rm{v_s}}\sin \vartheta\,(\cos(\omega_L t)\ex+\sin(\omega_L t)\ey)\!\cdot\!\svec$, induces time-dependent spin flips. 
Our model is a generalisation to arbitrary tunnelling coupling
of the one studied by Zhu and co-workers \cite{zhu2003,zhu2004}. 

We use the quasiclassical theory of superconductivity \cite{eilenberger1968,larkin1968,eliashberg1971,serene1983}
to solve the non-equilibrium tunnelling problem stated above. Within quasiclassical theory, interfaces are handled by the formulation of 
boundary conditions, which usually have been expressed as scattering problems
\cite{zaitsev1984,shelankov1984,millis1988,nagai1988,eschrig2000,shelankov2000,fogelstrom2000,barash2002,zhao2004,eschrig2009}.
In many problems, in particular when an explicit time dependence appears, we find
the t-matrix formulation more convenient to use\cite{cuevas1996,cuevas2001,andersson2002}.
This formulation is also well suited for studying interfaces with different numbers of trajectories on either side as is the case for normal metal/half metal
interfaces \cite{eschrig2003,kopu2004}. For a full account on how to solve the time-dependent boundary condition we refer to our original
articles \cite{teber2010,holmqvist2011,holmqvist2012,holmqvist2014}.

\begin{figure}[!b]
\centering\includegraphics[width=5.1in]{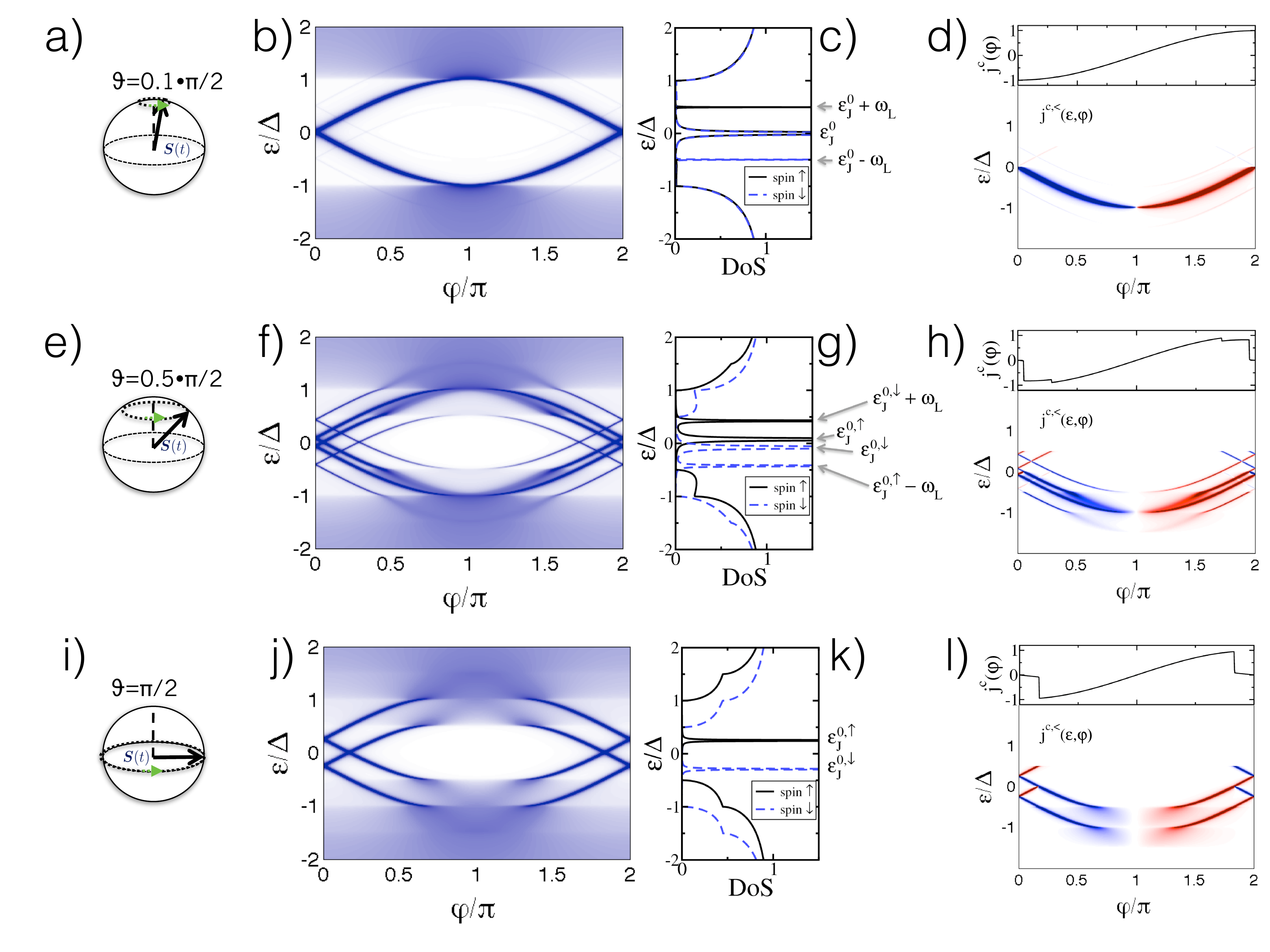}

\caption{The Josephson effect over a precessing spin at frequency $\omega_L=0.5\Delta$ for various cone angles (a-d) $\vartheta=0.1\pi/2$, 
(e-h) $\vartheta=\pi/4$, and (i-l) $\vartheta=\pi/2$. The tunnelling parameters are ${\rm {v_o}}=0, {\rm{v_s}}=1$ and the temperature is $T=10^{-5}\Delta$.
The structure of the Andreev-level spectrum is shown vs. phase in panels (b,f,j) \cite{holmqvist2010PhDthesis} and the density of states (DoS) at $ \varphi=0$ in (c,g,k) \cite{holmqvist2011}. The current-phase relations,
$j^c(\varphi)$, and the charge current kernels, $j^{c,<} (\varepsilon, \varphi)$, are shown in panels (d,h,l) \cite{holmqvist2011}. $j^{c,<} (\varepsilon, \varphi)$ shows how
the Andreev levels in (b,f,j) are populated and in which direction they carry current; red into the right and blue into the left lead. 
At some phase differences $\varphi_c<\varphi<2\pi-\varphi_c$, scattering between the Andreev levels and the continuum states broadens the otherwise
sharp in-gap states. The charge current (plotted in units of $e\Delta/\hbar$) is the energy-integrated spectral current and displays abrupt jumps at phase
differences where Andreev levels become populated/unpopulated. The DoS at $\varphi=0$ shows the splitting of the spin-up and spin-down
Andreev levels as well as the scattering of the continuum levels into the gap.  
}
\label{fig2}
\end{figure}

The quasiclassical propagator in lead $\alpha$, $\check{g}_\alpha$, is a $2\times2$ matrix in Keldysh space, denoted by the check "$\, \check{\,}\,$".
Each component is in turn a $4\times4$ matrix
in the combined Nambu-spin space and has the general form
\begin{equation}
\hat g^{R,A,K}=\left(\begin{array}{cc}g+\gvec\cdot\svec&(f+\fvec\cdot\svec)i\sy\\i\sy\,(\tilde f+\tilde \fvec\cdot\svec)&\sy(\tilde g-\tilde \gvec\cdot\svec)\sy
\end{array}\right)^{R,A,K}
\end{equation}
for the retarded ($R$), advanced ($A$), and Keldysh ($K$) components.
To obtain
$\check{g}_\alpha$ for a non-homogeneous system, we solve the transport equation 
\begin{eqnarray}
i v_F  \partial_x \check{g}_{\alpha}(\hat{p}_F) + [\check{\varepsilon}-\check{\Delta}_\alpha, \check{g}_{\alpha}(\hat{p}_F)  ]_\circ   
= \check{j}_{\alpha} \delta(x-x_c)/(2\pi i)
\end{eqnarray}
along a trajectory $\hat{p}_F$ in lead $\alpha$. 
The boundary conditions for the components of $\check{g}_\alpha$ enter via a localised inhomogeneity, given by the tunnel Hamiltonian, 
at the position of the contact, $x_c$ \cite{buchholtz1979,thuneberg1981,thuneberg1984}.
The source term is a matrix current defined as $\check{j}_{\alpha}/2\pi i= [\check{t}_{\alpha}(\hat{p}_F,\hat{p}_F), \check{g}^0_\alpha(\hat{p}_F)  ]_\circ$.
The $\circ$-product is a matrix multiplication and convolution over common time arguments and $\check{g}_\alpha$ additionally obeys a normalisation condition
$\check{g}_\alpha \circ \check{g}_\alpha=-\pi^2\check 1$. 
The matrix, $\check{t}_{\alpha}(\hat{p}_F,\hat{p}_F)$, solves the t-matrix equation
\begin{equation}
\check t_\alpha(t,t^\prime)=\check \Gamma_\alpha(t,t^\prime) 
+\lbrack \check \Gamma_\alpha\!\circ\!\check g^{0}_\alpha \!\circ\!  \check t_\alpha\rbrack (t,t^\prime).
\label{boundaryTmatrix}
\end{equation}
The t-matrix $\check{t}_{\alpha}$ depends on the hopping elements of Eq. \eqref{hopping element}
via a matrix $\check \Gamma_{L}(t,t^\prime)$ defined as
$\check \Gamma_{L}(t,t^\prime)=\lbrack \check v \!\circ\! \check g^{0}_{R}\!\circ\!\check v\rbrack (t,t^\prime)$
for the left side of the interface. The right-side matrix $\check \Gamma_{R}$ is correspondingly obtained from the left-side propagator $\check g^{0}_{L}$. $\check g^{0}_{L,R}$
are the bulk propagators in either lead computed without the tunnelling term.
From the t-matrices \eqref{boundaryTmatrix}, we calculate the full quasiclassical propagators, which can be separated into "incoming" ($\check g^{i}$) and "outgoing" ($\check g^{o}$) propagators depending on if their trajectories lead up to or away from the interface. These propagators are given by
\begin{equation}
\check g^{i,o}_\alpha(t,t^\prime)=\check g^{0}_\alpha(t,t^\prime)+
\lbrack (\check g^{0}_\alpha\pm i \pi \check 1)\!\circ\! \check{t}_\alpha \!\circ\!  (\check g^{0}_\alpha\mp i \pi \check 1)\rbrack(t,t^\prime),
\label{boundary}
\end{equation}
where $\pm$ and $\mp$ refer to the incoming and outgoing propagators, respectively. The matrix currents give the charge and spin currents via
\begin{subequations}
\begin{eqnarray}
j^{c}_\alpha(t) &=& \frac{e}{2\hbar} \int \frac{d \varepsilon}{8\pi i} \mbox{Tr} [ \hat{\tau}_3 
\hat{j}^{<}_{\alpha}(\varepsilon,t)]; \label{chargecurr} \\
\jvec^{s}_\alpha(t) &=&  \frac{1}{4}  \int \frac{d \varepsilon}{8 \pi i}  \mbox{Tr} [ \hat{\tau}_3 \hat{\svec} \hat{j}^{<}_{\alpha}(\varepsilon,t) ], \label{spincurr}
\end{eqnarray}
\end{subequations}
where $\hat \tau_3={\rm diag}(1,-1)$, $\hat{\svec}={\rm diag}(\svec,-\sy\svec\sy)$ and "$\, \hat{\,} \,$" denotes a $4\times 4$ matrix in Nambu-spin space. The 
lesser ("$<$") propagators can be obtained as 
$\hat{g}^<=(1/2)(\hat{g}^{K}-\hat{g}^{R}+\hat{g}^{A})$.
The itinerant electrons generate a spin transfer torque which gives a contribution to the torque in Eq. \eqref{LLG} as $\tvec=\jvec^s_L-\jvec^s_R$. 

The spin independence of $\check {g}^0_{\alpha}(\varepsilon)$ and the form of the hopping elements simplify the time-dependent problem.
This simplification can be made due to the fact that the Keldysh-Nambu-spin matrices can be factorised in spin space into generalised diagonal matrices, $\check{X}_d$, spin-raising matrices,
$\check{X}_\uparrow$, and spin-lowering matrices, $\check{X}_\downarrow$.
In general, a matrix factorised in this form has the time dependence
\begin{eqnarray}\label{Xmatrix}
\check X(t,t')=(2 \pi)^{-1}\int d\varepsilon \, {\rm e}^{-i\varepsilon(t-t')}\big\lbrack 
\check X_d(\varepsilon,\omega_L)+
+{\rm e}^{-i \omega_L t}\check X_\uparrow(\varepsilon,\omega_L)+
{\rm e}^{ i \omega_L t}  \check X_\downarrow(\varepsilon,\omega_L)\big\rbrack . 
\end{eqnarray}
The matrices $\check{X}_d$, $\check{X}_\uparrow$, and $\check{X}_\downarrow$ are still Keldysh-Nambu matrices and, in addition, obey the usual algebraic rules for spin matrices, i.e. $\check X_\uparrow\circ\check Y_\uparrow=\check X_\downarrow\circ\check Y_\downarrow=0$,
$\check X_{\downarrow,\uparrow}\circ\check Y_{\uparrow,\downarrow}\propto\check Z_{d}$, and 
$\check X_d\circ\check Y_{\uparrow,\downarrow}\propto\check Z_{\uparrow,\downarrow}$.
Observables, such as the charge and spin currents above, will have the general time dependence 
\begin{equation}\label{timedep}
{\cal{O}}(t;\omega_L)={\cal{O}}_o(\omega_L)+{\cal{O}}_z(\omega_L)\sz
+{\cal{O}}_\uparrow(\omega_L){\mathrm{e}}^{-i \omega_L t}\sigma_+
+{\cal{O}}_\downarrow(\omega_L){\mathrm{e}}^{i \omega_L t}\sigma_-.
\end{equation}
The components ${\cal{O}}_{o,z}$ are diagonal in spin space and have spin-angular
momentum $s_z=0$, while correspondingly ${\cal{O}}_{\uparrow,\downarrow}$ are off-diagonal in spin space and
have  spin-angular momentum $s_z=\pm1$.
In Eq. \eqref{timedep}, we have used the definitions $\sigma_{\pm}=(\sx\pm i\sy)/2$.

\section{Andreev-reflection-induced spin torques}\label{sec: AR torque}

Quasiparticle scattering in a Josephson junction may lead to the formation of Andreev levels if the scattering occurs in such that a way that the 
quasiparticles interfere constructively (see Fig.~\ref{fig1}(b)).
In the presence of a precessing spin, the quasiparticle scattering is modified by processes shown in Fig.~\ref{fig1}(c); a tunnelling quasiparticle may gain (lose)
energy $\omega_L$ 
while simultaneously flipping its spin from down (up) to up (down).
The Andreev level spectrum essentially depends on the ratio between the hopping amplitudes, ${\rm{v_o}}/{\rm{v_s}}$. If ${\rm{v_o}}/{\rm{v_s}}<(>)1$, the 
junction is in a $\pi(0)$ state \cite{holmqvist2011,teber2010}, see Fig.~\ref{fig1}(d).
The additional precession-induced tunnelling processes modify the Andreev levels. The Larmor frequency, $\omega_L$, determines the amount of energy 
exchanged during a tunnelling event, while the cone angle, $\vartheta$, determines the amount of scattering between the spin-up and -down bands.
These parameters, as well as the temperature, determine the population of the Andreev states \cite{holmqvist2011,holmqvist2012}.
In Figure \ref{fig2}, we summarise how the tunnelling over a precessing spin modifies the Andreev spectra by introducing scattering resonances
created by the combination of quanta exchange of $\hbar\omega_L$ and spin flips.
The charge current is time-independent but still dependent on both $\omega_L$ and $\vartheta$ as seen in
Fig.~\ref{fig2}.
While the Josephson effect over the precessing spin is interesting in its own right, 
we will not discuss the current-phase relations further in this paper and refer the interested reader to the original articles 
\cite{teber2010,holmqvist2011,holmqvist2012}. Instead, we will focus on the effects of dynamic spin-triplet correlations and their
consequences.  

An s-wave superconductor contains only spin-singlet correlations 
$\sim \frac{1}{2}\langle\psi_\uparrow\psi_\downarrow-\psi_\downarrow\psi_\uparrow\rangle$ and can not support a spin current.
Nevertheless, induced spin-triplet correlations can be formed due to spin mixing and locally broken spin-rotation 
symmetry \cite{eschrig2008,houzet2008,alidoust2010}.
The rotation of the classical spin generates new spinful correlations and spin currents that are created by 
the Andreev processes depicted in figure \ref{fig1}(b-c); positive interference along closed loops leads to the spin-triplet correlations
$\frac{1}{2}\langle\psi_\uparrow\psi_\downarrow+\psi_\downarrow\psi_\uparrow\rangle$, 
$\langle\psi_\uparrow\psi_\uparrow\rangle$ and $\langle\psi_\downarrow\psi_\downarrow\rangle$.
These correlations depend on the characteristics of the tunnelling interface, i.e. the precession frequency, $\omega_L$,
the cone angle, $\vartheta$, the relative amplitude of hopping strengths, ${\rm{v_o}},{\rm{v_s}}$, as well as the 
superconducting phase difference $\varphi$, and the temperature, $T$.
These spin-triplet correlations are localised near the junction interface and decay over length scales 
on the order of the superconducting coherence length \cite{fogelstrom2000,shevtsov2014}.

The spin-singlet components can be quantified by
$\psi(\hat{k})=\int_{-\varepsilon_c}^{\varepsilon_c} d \varepsilon \lbrack f^<(\hat k,\varepsilon)+f^<(-\hat k,\varepsilon) \rbrack / 8 \pi i$,
where $f^<(\pm \hat{k},\varepsilon)$ denotes the anomalous Green's functions at the Fermi-surface points $\pm \hat{k}$.
$\psi(\hat{k})$ is a measure of the (singlet) pairing correlations available to form a singlet order parameter 
$\Delta_s(\hat{k})= \lambda_s \eta(\hat{k}) \langle  \eta(\hat{k}^\prime) \psi(\hat{k}^\prime)\rangle_{\hat{k}^\prime\cdot \hat{n}>0}$, where 
$\eta(\hat{k})=\eta(-\hat{k})$ are basis functions of even parity on which the pairing interaction may be expanded and $\hat{n}$ is the direction of the surface normal. 
The energy $\varepsilon_c$ is the usual cut-off that appears in the BCS gap equation. 
The triplet correlations span the spin space in 
such a way that $\fvec_z^<\sim\frac{1}{2}\langle \psi_\uparrow \psi_\downarrow+ \psi_\downarrow\psi_\uparrow 
\rangle$ and $\fvec^<_{\uparrow/\downarrow}\sim \langle \psi_{\uparrow/\downarrow} \psi_{\uparrow/\downarrow} \rangle$. 
We quantify the induced spin-triplet correlations, $\fvec^{<}$, in terms of a $\dvec$ vector, which in general is a $2\times2$ triplet order parameter 
given by $\Delta_{\hat{k}}=\dvec(\hat k)\!\cdot\!\svec\, i \sigma_y$ and points along the direction of zero spin projection of the Cooper pairs \cite{he3book}. 
We make the following definitions:
\begin{subequations}\label{triplet correlations}
\begin{eqnarray}
\pi \,\,{\rm junctions} &\quad &\dvec_{o}(\hat k)= \hat n\!\cdot\! \hat k \int_{-
\varepsilon_c}^{\varepsilon_c} \frac{d \varepsilon}{8 \pi i} 
\lbrack \fvec^< (\hat k,\varepsilon)-\fvec^< (-\hat k,\varepsilon) \rbrack ,\\
0 \,\,{\rm junctions} &\quad & \dvec_{ e}(\hat k)= \int_{-\varepsilon_c}^{\varepsilon_c} \frac{d \varepsilon}{8 \pi i} s_\varepsilon \lbrack \fvec^< (\hat k,\varepsilon)+ \fvec^< (-\hat k,\varepsilon) \rbrack ,
\end{eqnarray}
\end{subequations}
where the vector $\dvec_o$ is odd in momentum and even in energy, and the vector $\dvec_e$ is even in momentum and odd in energy.
$s_\varepsilon$ is the sign of the energy $\varepsilon$.
Spin-triplet pairing that is {\em even-in $\hat{k}$} and {\em odd-in $\varepsilon$} was first considered as a candidate pairing state for $^3$He \cite{berezinskii1974} and has recently been realised in superconductor/inhomogeneous magnet interfaces \cite{dibernardo2015}.
The time-dependence of the $\dvec$ vector follows from Eqs. \eqref{Xmatrix}, \eqref{timedep}, i.e.
\begin{equation}
\dvec(t)= \dvec_z + \dvec_\uparrow {\rm e}^{-i \omega_Lt}+\dvec_\downarrow {\rm e}^{i \omega_Lt}.
\end{equation}
For ${\rm v_o}=0$ and finite ${\rm{v_s}}$, the components are equal in magnitude, $\dvec_\uparrow=\dvec_\downarrow=-\dvec_z$ and scale with a 
common prefactor, ${\cal{D}}_{s}\omega_L$, where ${\cal{D}}_{s}=4 {\rm{v_s}}^2/[1+2({\rm{{\rm v_o}^2+v_s}}^2)+({\rm v_o}^2-{\rm{v_s}}^2)^2]$.
As expected, the $\dvec$-vector components decrease for increasing temperature until they vanish at $T=T_c$.
For finite values of ${\rm v_o}$, the universal scaling disappears and the $\dvec$-vector components display an asymmetry between 
$\dvec_{\uparrow}$ and $\dvec_{\downarrow}$.
For temperatures $T/T_c\lesssim0.1$, the $\dvec$ vector can be expressed in terms of the classical spin,
\begin{equation}
\dvec(t)=\delta_L \dot\SMM(t)\!\times\!\SMM(t)+\delta_H (\gamma \Heff)\!\times\!\SMM(t)+\delta_z \SMM_z.
\label{deltadvec}
\end{equation}
For the odd $\dvec$ vector, $\delta_{z,o}=0$ and, in the tunnel limit at zero temperature, $\delta_{L,o} = \pi{\cal{D}}_{s}\sin(\varphi/2)$ and 
$\delta_{H,o} = 4\pi i {\rm{v_o}}{\rm{v_s}} \sin(\varphi/2)$.
The $\dvec$ vectors in the left and right leads are related by $\dvec_{R}(t)=-\dvec_L(t)$.

The spin-vector part of the normal Green's function, $\gvec^{R/A}$, can be expressed in terms of the spin-vector part of 
the anomalous Green's functions, $\fvec^{R/A}$, using the normalisation condition. In the limit of a small cone angle, the $z$ component is negligible and
\begin{eqnarray}
g^{R(A)}_{\uparrow/\downarrow,\alpha}\left( \varepsilon \mp \frac{\omega_L}{2} \right) = \frac{1}{\bar{g}^{+/-,R(A)}_{s,\alpha}} \bigg \{ \bigg [ \frac{(1\pm i)}{2}f^{R(A)}_{s,\alpha}\left(\varepsilon\pm \frac{\omega_L}{2}\right) + \frac{(1\mp i)}{2}f^{R(A)}_{s,\alpha}\left(\varepsilon\right) \bigg ] \tilde{f}^{R(A)}_{\uparrow/\downarrow,\alpha}\left( \varepsilon   \right)  \nonumber \\
+   \bigg [ \frac{(1\mp i)}{2}\tilde{f}^{R(A)}_{s,\alpha}\left(\varepsilon\pm \frac{\omega_L}{2}\right) + \frac{(1\pm i)}{2}\tilde{f}^{R(A)}_{s,\alpha}\left(\varepsilon\right) \bigg ] f^{R(A)}_{\uparrow/\downarrow,\alpha}\left( \varepsilon  \right)    \bigg \} ,
\end{eqnarray}
where $\bar{g}^{+/-,R(A)}_{s,\alpha}=g^{R(A)}_{s,\alpha}\left( \varepsilon \right) + g^{R(A)}_{s,\alpha}\left( \varepsilon\pm \omega_L/2 \right)$ and $g^{R(A)}_{x,
\alpha}=[g^{R(A)}_{\uparrow,\alpha}+g^{R(A)}_{\downarrow,\alpha}]/2$ and $g^{R(A)}_{y,\alpha}=i[g^{R(A)}_{\uparrow,\alpha}-g^{R(A)}_{\downarrow,\alpha}]/2$.

It is then clear that 
the existence of
the spin currents, 
$\jvec^s_\alpha=(1/4)\int (d\varepsilon/8 \pi i) {\rm Tr} \{ \hat{\tau}_3 \hat{\svec} [ \hat{g}^{i,K}_{\alpha}(\varepsilon,t)- 
\hat{g}^{o,K}_{\alpha }(\varepsilon,t) ] \}$,
are a direct consequence of the precession-induced spin-triplet correlations. See also Appendix in Ref.~\cite{holmqvist2012}.
Unfortunately, the spin currents decay over relatively short distances, viz. the superconducting coherence length, and are therefore difficult to measure.
The spin current is nothing but transport of spin-angular momentum and the non-conservation of the spin current results in a torque acting on the rotating spin
thereby creating a back-action on the precessing spin that is sufficiently large for experimental detection \cite{holmqvist2011}, as will be described below.

Since $\jsvec^{s}_{R}(t,\varphi)=-\jsvec^{s}_{L}(t,-\varphi)\neq \jsvec^{s}_{L}(t,\varphi)$, the difference between the spin currents can be used to calculate the torque $\tauvec(t)$ in Eq. \eqref{LLG}.
We call this torque the Andreev torque since it has its origin in the Andreev scattering processes described in Fig.~\ref{fig1}. The torque contribution per conduction channel is
\begin{eqnarray}
\tauvec_A(t)=\frac{2\hbar}{S}{\cal{D}}_{s}\beta_H\cos\vartheta \, (\gamma\Heff)\!\times\! \SMM(t).
\label{torque A}
\end{eqnarray}
This torque describes a shift of the precession frequency, $\omega_L\rightarrow \omega_L\lbrack 1+\frac{2\hbar}{S}{\cal{D}}_{s}\beta_H \cos\vartheta\rbrack$,
and this shift is therefore a direct consequence of the induced spin-triplet correlations.

A measurement of this frequency shift is a measurement of the induced spin-triplet correlations. Since the shift is $\propto 1/S$, we suggest a nanomagnet with a spin that is small, but still large enough to be treated as a classical spin, say a magnetic nanoparticle with spin $S\sim50\hbar$.
For a contact with two superconducting
niobium (Nb) leads, the effective contact area is $\sim\pi\xi_0^2$, where the superconducting coherence length
$\xi_0\sim40\,$nm for Nb.
A contact width of $\sim40$ nm contains $n\sim200$ conduction channels. In bulk Nb, $\Delta\sim1\,$meV, but can be made considerably smaller in the point contact, say $\Delta\sim200\,\mu$eV.
We can now study the changes to the precession due to the Andreev torque.
In a typical FMR experiment, the resonance peak in the power absorption spectrum has a width that is produced by inhomogeneous broadening, e.g. from anisotropy fields, and homogeneous broadening, which is due to Gilbert 
damping, and can be expressed as $\Delta H_{\rm hom}=\frac{2}{\sqrt{3}} H \alpha_{G}$ \cite{platow1998}, where $H=\vert \Heff \vert$ and $\alpha_G=\frac{2\hbar}{S} n \alpha {\cal{D}}_{s}$ is the Gilbert constant \cite{gilbert2004}.
A typical magnetic field is $H\sim 180$ mT, which corresponds to a Larmor precession of $\sim20\,\mu$eV or $5$ GHz. Here, we have assumed a uniform precessional motion.
In Ref.~\cite{holmqvist2011}, it was shown that the normal quasiparticles freeze out as the temperature is lowered. This process results in a decrease of the width of the resonance peak \cite{bell2008}. For a junction with ${\cal{D}}_{s}\sim 0.1$, the difference in homogeneous broadening is on the order of
$\Delta H_{\rm hom}(T/T_c>1)- \Delta H_{\rm hom}(T/T_c\rightarrow 0) \sim 80$ mT.
In addition to the resonance peak width reduction, the shift of the resonance peak $H_0$ due to the Andreev torque appears.
The frequency shift corresponds to $\Delta \omega_L / \omega_L=\alpha_G \beta_H \cos\vartheta$. In the tunnel limit, $\beta_H\sim\frac{1}{16}\frac{\omega_L}{\Delta}$ in the low temperature limit \cite{holmqvist2011}.
In this limit, a spin with angle $\vartheta=\pi/4$ can hence generate
a displacement of the resonance peak by $\Delta H_0/H_0\sim 2\%$.
By increasing the junction transparency, the ratio $\hbar n/S$, or the ratio $\omega_L/\Delta$, the ratio $\Delta H_0/H_0$ can be improved.

\section{Spin-precession assisted multiple Andreev reflection}

\begin{figure}[!t]
\centering\includegraphics[width=4.9in]{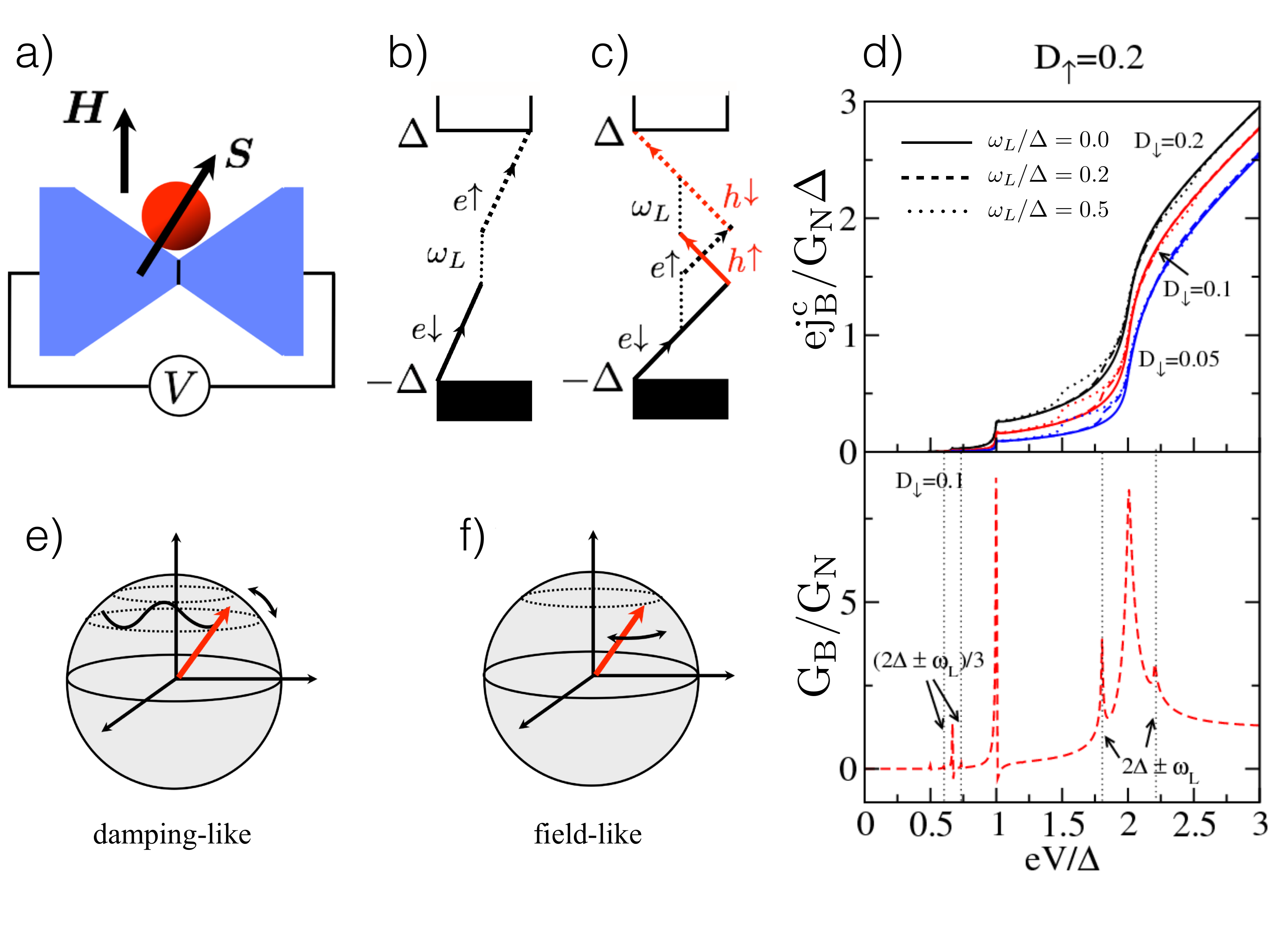}
\vspace{-0.7cm}
\caption{(a) Same setup as in Fig.~\ref{fig1} but with a bias voltage applied across the tunnel junction.
The (b) first- and (c) second-order MAR processes combined with absorption of energy $\omega_L$.
(d) Current-voltage characteristics for the dc charge current $j^c_B$ (top)
and differential conductance (bottom), $G_B=\partial j^{c}_{\rm{B}}/\partial V$, normalised by the normal conductance $G_N=[e^2/h][{\cal D}_\uparrow+{\cal D}_\downarrow]$. In both plots, $\vartheta=\pi /8$.
Sketches of the time-dependent (e) damping-like and (f) field-like torques created by spin-precession-assisted MAR.
}
\label{fig3}
\end{figure}

Replacing the phase bias by a voltage bias (Fig.~\ref{fig3}(a)) leads to several new features \cite{holmqvist2014} attributable to the interplay 
between the 
time-dependent $\dvec$ vectors and the Josephson frequency, $\omega_J=2eV/\hbar$. The replacement causes the phase difference to increase linearly 
in time, $\varphi(t)=\varphi_0+\omega_J t$, where $\varphi_0$ is the initial phase difference.
The bias voltage in combination with energy exchange with the precessing spin creates multiple Andreev reflection (MAR) processes that lead to characteristic signatures in the charge current-voltage characteristics \cite{octavio1983,bratus1995,averin1995,cuevas1996}.
Two examples of spin-precession-assisted MAR are shown in Fig. \ref{fig3}. 
Similarly to the phase-biased case, energy absorption (emission) corresponds to spin flip from down (up) to up (down).
The first-order process shown in Fig.~\ref{fig3}(b), which includes an energy absorption of $\omega_L$, leads to a contribution to the IV characteristics at the
 energy $eV=2 \Delta-\omega_L$. Fig.~\ref{fig3}(c) shows the two possible second-order processes that include absorption of energy.
The spin flip associated with the energy exchange introduces a minus sign in the next Andreev-reflection amplitude due to the change between the spinors
$(\psi_\uparrow,\psi^\dagger_\downarrow)^T \leftrightarrow (\psi_\downarrow,\psi^\dagger_\uparrow)^T$.
This sign difference leads to destructive interference and suppression of the total Andreev reflection. Destructive interference occurs for all even processes,
$n=2,4,...$, while higher-order odd processes display constructive interference.

The bias voltage makes the calculations of the charge and spin currents considerably more complicated. This complication arises in large due to the MAR 
processes, which make it impossible to express the Green's functions using a closed set of equations. Instead, a recursive approach, see 
Ref.~\cite{holmqvist2014} for details, has to be used.
The general time dependence of a general matrix such as $\check X(t,t')$ in Eq. \eqref{Xmatrix} now has to be complemented by the time dependence 
generated by the Josephson frequency.
In general, the current is given by
\begin{equation}\label{eq:jmu}
j_\alpha^\mu(t) =  \sum_{n,m} e^{-i(n\varphi_0+m\chi_0)-i(n\omega_J+m\omega_L)t} (j^\mu_\alpha)^m_{n}\, .
\end{equation}
The current components are
\begin{equation}\label{eq: j n m component}
(j^\mu_\alpha)^m_{n} =  \int \frac{d\varepsilon}{4}{\rm Tr}\{\hat{\kappa}^\mu[  \check{t}(\varepsilon+n\omega_J+m\omega_L) \check{g}(\varepsilon)-
\check{g}(\varepsilon+n\omega_J+m\omega_L)  \check{t}(\varepsilon+n\omega_J+m\omega_L) ]^< \} ,
\end{equation}
where we have defined $\hat\kappa^0=e\hat\tau_3$ for the charge current and $\hat{\kappa}^i={\rm diag} (\sigma_i, \sigma_y\sigma_i\sigma_y)/2$ 
for a spin current with a polarisation in the $i=x,y,z$ direction.  
Note that just as the current depends on the initial phase $\varphi_0$, it also depends on $\chi_0$, which is the initial value of the in-plane projection of 
the precessing spin.
The integer $m$ takes the values $\{-1,0,1\}$ corresponding to $\{\downarrow,d,\uparrow\}$ in Eq. \eqref{Xmatrix}. 
Defining ${\rm v}_{\uparrow/\downarrow}={\rm v_o} \pm {\rm v_s} \cos\vartheta$, we write 
${\cal D}_{\uparrow(\downarrow)}=4{\rm v}_{\uparrow(\downarrow)}^2/[1+{\rm v}_{\uparrow(\downarrow)}^2]^2$.

The dc charge current and the differential conductance, plotted in Fig.~\ref{fig3}(d), clearly show the contributions to the current generated by the 
spin-precession-assisted MAR processes.
These features appear at voltages $eV=(2\Delta\pm\omega_L)/n$, where $n=1,3,...$
Note that, as expected, the contributions for the even processes $n=2,4,...$ are absent.
It can be shown that the ac charge current only includes harmonics of 
$\omega_J$, i.e. $j_\alpha^0(t) =  \sum_{n} e^{-in(\varphi_0+\omega_J t)} (j^0_\alpha)^0_{n}$. 
This time dependence is an effect of the combined energy exchange-spin flip tunnelling processes.

The spin current, on the other hand, includes all harmonics of the Larmor and Josephson frequencies. This time dependence is 
captured by the spin-transfer torque, whose $\omega_L$ dependence
is described by the expression
\begin{equation}\label{eVtorque}
\tauvec(t) = \frac{\gamma_H(t)}{S}  \gamma \Heff \times \SMM +\frac{\gamma_L(t)}{S^2}  \dot{\SMM} \times \SMM ,
\end{equation}
where the prefactors, $\gamma_{H/L}$, oscillate with the Josephson frequency, $\gamma_{H/L}(t)= \sum_{n} \gamma_{H/L,n}e^{i n \omega_J t}$.
The component $\gamma_{L,0}$ describes a finite shift of the precession angle $\vartheta$, while the term $\propto \gamma_{H,0}$ signals a shift 
of the precession frequency. The damping-like torque $\propto \gamma_{L,n}$ and the field-like torque $\propto \gamma_{H,n}$ describe 
Josephson nutations \cite{nussinov2005} and oscillations of the precession frequency, respectively.

Since the torque \eqref{eVtorque} includes harmonics of both $\omega_J$ and $\omega_L$, resonances may occur when the two frequencies are 
commensurate. These Shapiro resonances occur at the bias voltage $V_n^m=-(m/n)\omega_L/2e$ where $n,m \neq 0$, and results in a dc contribution 
to the spin-transfer torque and can be seen as a rectification of the higher harmonics of the torque in section \ref{sec: AR torque}.
As the ac part of the torque \eqref{eVtorque} originates from an in-plane spin-polarised current, 
one can then conclude that the Shapiro resonances produce dc in-plane torque components.
The Shapiro resonances hence break the rotational symmetry around the $z$ axis and, therefore, the Shapiro 
torque depends on the initial angle of nanomagnet's magnetisation direction, $\chi_0$. This situation is analogous to 
the $\varphi_0$-dependence for the Shapiro steps seen in microwave-irradiated Josephson junctions \cite{cuevas2002,uzawa2005,chauvin2006}.

The dc Shapiro torque will cause the spin to precess around a new $z$ axis.
Choosing suitable parameters and applying a self-consistent solution, one finds that the Shapiro torque is able to reverse the spin's direction. 
To this end, we choose $n=1$ and optimise the effect of the Shapiro torque by maximising the ratio $\gamma_{H,1}/\gamma_{L,1}$. It was found in 
Ref.~\cite{holmqvist2014} that $\gamma_{H,1}$ strongly depends on the junction transparency but exhibits a weak dependence on the precession 
angle. We therefore choose $\rm{v_o}=0$, $\rm{v_s}=0.7$, and $\vartheta=0.1 \pi$. We consider a tunnel junction consisting of Nb having a superconducting gap $\Delta\sim 0.5$ meV and containing a magnetic nanoparticle with spin $S\sim50\hbar$ with a typical frequency $\omega_L\sim 5$ GHz that corresponds to a magnetic field well below the critical magnetic field.
We therefore have $\omega_L/\Delta=0.01$. A magnetic field close to the critical magnetic 
field reduces $\Delta$ and increases the resolution of features depending on the ratio $\omega_L/\Delta$, e.g. the subgap features in the dc charge current.
A point contact of width $\sim 40$ nm has $\sim200$ conduction channels, which gives an estimated sub-nanosecond switching time for the first 
Shapiro resonance.

\section{Conclusion}
We have reviewed recent work on how the magnetisation dynamics of a nanomagnet couple to the charge and spin Josephson effects.
The precession of the nanomagnet modifies the Andreev scattering in several ways. First, it introduces a spin-polarised Andreev level spectrum
and dynamical spin-triplet pairing correlations in the vicinity of the junction. Second, it couples in-gap Andreev levels with the
continuum part of the spectrum causing a nonequilibrium population of the Andreev levels.
Third, it creates a nonequilibrium population of the Andreev levels, leading to Andreev levels carrying current in opposite directions being populated and a strongly modified current-phase relation.
We have focused on the consequences of the spin-polarised Andreev-level 
spectra and how they couple back to the precession dynamics of the nanomagnet via conservation of spin-angular momentum.
Depending on if the Josephson junction is phase biased
or voltage biased, the this torque can modify the precession frequency, either by a frequency shift or by frequency modulations, or it can
introduce nutations. Recent experiments on superconductor/ferromagnet nanojunctions can extract the microscopic details of the scattering and 
match junction parameters
such as spin-filtering and spin-mixing effects \cite{quay2013,hubler2012,wolf2013}. If the ferromagnetic part of the junction would be a single domain magnetic
grain, properties described in this review could be probed in experiments.   
\vskip6pt

\ack{C.H. and W.B. were supported by the DFG and SFB 767. M.F. acknowledges support from the Swedish Research Council (VR).}
\enlargethispage{20pt}


\end{document}